# Adaptive Traffic Signal Control with Deep Reinforcement Learning – An Exploratory Investigation

**Matthew Muresan**

Ph. D. Candidate, Department of Civil & Environmental Engineering,
University of Waterloo,
Waterloo, ON, N2L 3G1, Canada
Email: mimuresa@uwaterloo.ca

**Liping Fu**\*

Professor, Department of Civil & Environmental Engineering, University of Waterloo, Waterloo, ON, Canada, N2L 3G1; Intelligent Transportation Systems Research Center, Wuhan University of Technology, Mailbox 125, No. 1040 Heping Road, Wuhan, Hubei 430063
Email: lfu@uwaterloo.ca

**Guangyuan Pan**

Postdoc Fellow, Department of Civil & Environmental Engineering,
University of Waterloo,
Waterloo, ON, N2L 3G1, Canada
Email: g5pan@uwaterloo.ca

**ABSTRACT**

This paper presents the results of a new deep learning model for traffic signal control. In this model, a novel state space approach is proposed to capture the main attributes of the control environment and the underlying temporal traffic movement patterns, including time of day, day of the week, signal status, and queue lengths. The performance of the model was examined over nine weeks of simulated data on a single intersection and compared to a semi-actuated and fixed time traffic controller. The simulation analysis shows an average delay reductions of 32% when compared to actuated control and 37% when compared to fixed time control. The results highlight the potential for deep reinforcement learning as a signal control optimization method.

**INTRODUCTION AND BACKGROUND**

The growth of cities brings new economic opportunity; however, it also brings with a number of challenges, including an increased burden on infrastructure and an increase in demand using them, leading to traffic congestion. A study done in 2013 indicated that, on average, road uses spend 36 hours per hear in traffic in cities worldwide, and by 2030 is expected to increase to 43 hours. In the United States, these effects are estimated to cost $78 billion in additional fuel costs and wasted time to users, and an additional $45 billion in wider economic effects (*1*). These congestion problems are the norm in many cities worldwide, and are consistently considered as one of the most significant problems facing urban centres as they grow. With continued growth expected, there is a strong impetus to find solutions that can mitigate these negative effects through traffic management strategies. One of the most visible and important infrastructures in this regard are traffic signals.

With the advent of the so-called "Big Data Era," systems worldwide now have access to a variety of rich data sources, and more are expected to become available as connected vehicle technologies increasingly become available. Despite these changes, the majority of traffic control systems are still dependent on legacy technologies and methods. In the United States alone, nearly 95% of traffic signals are operated using traditional methods developed before this Big Data Era (*2*). These technologies are unable to respond to changing conditions without re-timing from maintenance personnel, and various studies have citied reductions in delay of nearly 10% from simple signal re-timing (*3*).

Beyond data availability, high profile advances in Machine Learning have highlighted the capability of computer systems to replicate human behavior and best even the most highly trained individuals in an area. One such example is the recent defeat of the world's best Go player, Ke Jie, by AlphaGo. The game of Go is a traditional board game of high complexity, and until this time computers had not been able to defeat the best human players. By using a deep artificial neural network, AlphaGo was trained using a variety of techniques, including using reinforcement learning to play games against itself (*4, 5*).

Novel applications of these techniques continue to made across different fields. While today's traffic systems may still be dependent on legacy data sources and technologies, this represents an opportunity to introduce and develop the next generation of intelligent traffic signal systems. The objective of this study was to investigate the potential of applying deep learning models to control traffic signals. Applications of deep learning to signal control is a relatively new field, and many unanswered questions remain. This paper explores some aspects of these systems and discusses the results of a novel deep reinforcement algorithm for signal control. The remainder of this paper reviews some literature related to these topics, discusses the methodology applied, and goes over the results of the analysis conducted.

**Literature Review**

*Existing Traffic Control Systems*

Traffic signal control strategies can be roughly divided into three families: fixed time systems, actuated systems, and adaptive systems. Systems in the first two form the majority of deployed systems worldwide, with systems in the third category often seen only in large cities. For fixed time and actuated time signals, timing plans are normally generated through the use of software such as Synchro or TRANSYT. These methods solve for the optimal signal timing plan using a performance measures derived from intersection traffic models, such as delay, signal progression, stops, or degree of saturation. These software packages then use hill-climbing, genetic search, or some other method to find optimum combinations of cycle length, green time and offset lengths (*6, 7*). Actuated systems extend this by adding vehicle detection on some or all of the approaches. A variety of configurations can be used, including using vehicle detectors to serve minor approaches when vehicles are detected, or using them to allow a signal to terminate early (called "gapping out") to serve a conflicting movement (*8*). The dependence on the pre-configured signal timing plan, however, means that these systems still need timing updates as the city develops. Furthermore, semi-actuated configurations with detectors on the minor approach only are the most common implementation, and so in practice these systems operate similarly to fixed time configurations.

To address the limitations of fixed time systems, a variety of adaptive signal control strategies have also been developed. Among these the Split Cycle Offset Optimization Technique (SCOOT) and the Sydney Coordinated Adaptive Traffic System (SCATS) have seen the most deployment. As a centralized system, SCOOT is designed operate on a network of loop detector systems. Typically, detectors are normally required on every link upstream of an intersection controlled by the system, as the computer controlling the intersection uses this information to build flow profile estimates and predict whether changes in the split time, offset, and cycle length would be optimal (*9, 10*). Similarly, SCATS also optimizes the same set of parameters and also requires detectors on each intersection. Unlike SCOOT, it is a hierarchical system, and so optimization of certain components is divided between the different model levels.

To address the limitations in other popular adaptive systems, new systems are being developed and tested. One such system is RHODES, which uses a dynamic network loading model that captures the time-varying nature of traffic and has been field tested on a limited scale. RHODES makes decisions to support traffic flow through the traffic network by estimating specific travel flow characteristics, such as the speeds that platoons of vehicles move at, and the queues they will create. The system is more complex than SCATS or SCOOT, and consequently needs uses upstream detectors placed just after a traffic light (*11*). Field deployment of the model was initially conducted at two intersections of a highway and arterial diamond interchange. The resulting evaluation showed that the performance of the system matched a well timed semi-actuated controller in most situations (*12*).

Development of advanced adaptive systems is an area that has seen strong research, and have been encouraged by campaigns and programs such as the FHWA's *Every Day Counts* initiative that promotes the use of innovative technologies such as adaptive signal control. New proprietary systems have been developed in response to these pushes that are designed to use different data sources, such as video feeds. One such system is InSync, which is a proprietary system developed by Rhythm Industries. This system is designed to process real-time images of traffic and extract information on demand levels, which are in turn used to determine which movements receive green and for how long (*13, 14*). The system has already seen success and is deployed in over 30 states and 1 Canadian province, and highlights the opportunity for new technologies to play a role in traffic signal control.

*Machine Learning and Signal Control*

Instead of using traditional approaches to adaptive control that require a model of the traffic system, recently researchers have been exploring machine learning approaches to signal control. For example, El-Tantawy et al proposed a multi-agent control system based on reinforcement learning (RL) and tested it on a simulated representation of the City of Toronto (*15*).  In a RL based traffic control system, an agent is designed as the traffic controller to interact with the environment (intersection and traffic), and based on the actions it takes (change signal indications) it receives a reward or penalty. The agent is normally given a finite representation of this environment in the form of *states*, which may be the locations of vehicles on the approaches. Rewards are defined by the modeler, such as rewarding the agent for each vehicle it discharges, or penalizing the agent for each vehicle it delays, and the agent's goal is to seek an action policy that maximizes these rewards. In the system proposed by El-Tantawy et al., the states were characterized in terms of queue lengths and the current phases that were being timed. In its cooperative mode, the system state also included the states of neighboring intersections. Their system was rewarded based on the difference in delay between two successive decision points, with reductions in delay corresponding to positive reward. Through testing of their system on a simulated subset of Toronto's network, they found that intersection delays were reduced by 27% when controllers did not cooperate and 39% when controllers were set in cooperative mode. Subsequent study has also been conducted on the use of RL for signal control with promising results, such as a study by Balaji et al. that showed reductions of 15% compared to conventional systems (*16*), and a study by Wiering that also showed substantial reductions (*17*).

To determine the optimal policy, in RL an action-value function called "Q" is typically used. The process is typically referred to as "Q-learning", and is based on the form first proposed by Watkins (1989), which usually takes the following form (*18*):

$$Q(s_t, a_i) = r(s_t, a_i) + \gamma \max Q(s_{t+1}, a_i)$$

Where *r* is the expected immediate reward given state *s* at time *t* if action $a_i$ is taken, and γ is a discount factor applied to expected rewards from all future states. Typically an iterative process is used to build the values of this function, with the expected reward for choosing a state slowly modified based on the agents experience.

Recently, improvements in performance have been reported by researchers in using artificial neural networks to approximate this function. For example, a study by Li et al. compared a deep learning-based approach to approximate the Q function to regular reinforcement learning and found that the results produced delay reductions of about 14% for identically framed states (*19*). Another study by Genders and Razavi compared a deep structure to a one-layer neural network and found significant improvements (*20*). As an emerging field, however, many unanswered questions still remain. In particular, the optimal representations of state space, actions and rewards is still an unsettled question.

**MODEL DEVELOPMENT**

To analyze the optimal structure for a deep-learning based signal control strategy, an integrated framework is being developed with the Python Programming Language, allowing easy use of Google's TensorFlow as the core structure. TensorFlow, developed at Google by various researchers (*21*), is an open source interface that can be used to implement and execute many machine learning structures. It provides Python modules that can be used to permit easy use of TensorFlow's powerful mechanics. By connecting this with a vehicle simulator, an integrated framework can be created. This research uses VISSIM for this purpose, which is a microscopic simulator that implements a psycho-physical car following model (*22*). Like TensorFlow, VISSIM provides an interface that can be accessed in Python, allowing information about the position, speed, and route choice behavior of vehicles to be extracted. Within this framework, we propose a deep reinforcement learning with a novel space state characterization for use in optimal signal control. The structure of the model and interaction of the various components is presented in Figure 1. In this paper, we examine its effectiveness for the case of a single intersection as is trained over multiple days. The remaining sections of this paper highlight the various components of the model we propose, and some results from preliminary evaluation on a single intersection.

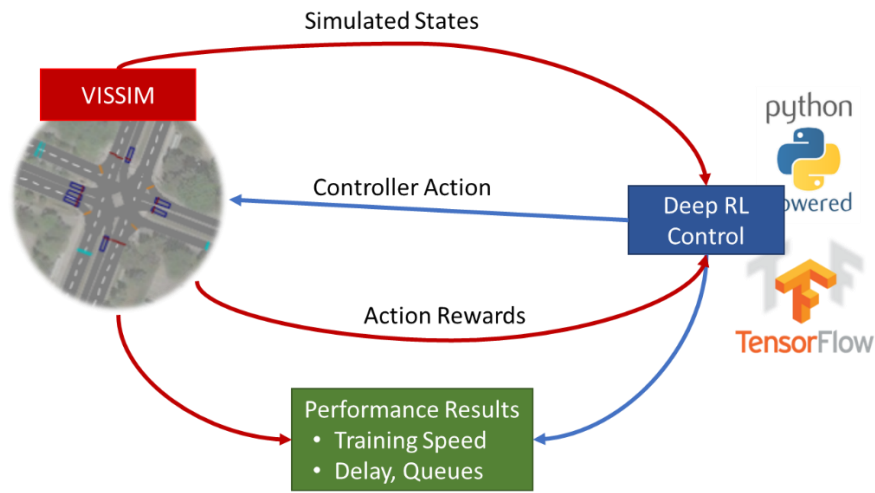

**FIGURE 1 – Deep Reinforcement Learning Based Traffic Control - Model Structure**

**Simulation Assumptions**

The environment simulated in this case was for a single intersection with no turning movements. This was chosen for simplicity in this evaluation, and the proposed system has been designed to control more complex situations. Volumes on each approach are time-varying, as shown in Figure 2. For simplicity, average volumes are held constant in VISSIM for at least 1 hour, but as arrivals are stochastic, the exact arrival pattern will fluctuate. This is done due to the method used to code the volumes in VISSIM, which allows specification of the average volume for a time period.. The analysis discussed in the results section also includes an "unusual event" scenario where an increase in volume on the minor street during the hours of 20:00 and 21:00 is tested.

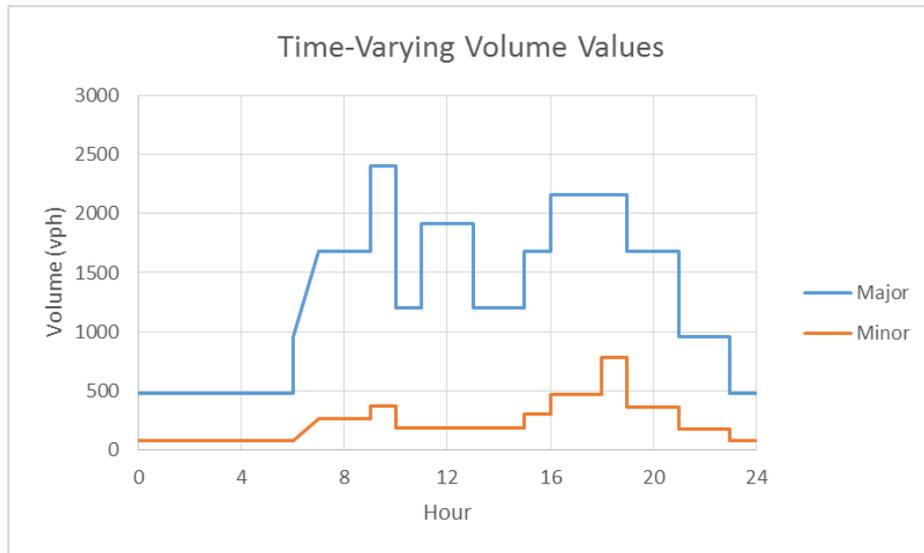

**FIGURE 2 – Time-Varying Volume**

**Deep Reinforcement Model**

Configuration of a deep reinforcement learning model requires specification of three important aspects: the state representation, the reward function, the model structure, and the action space definition. The following sections highlight the design decisions made in this regard.

*State Space Representation*

Many previously proposed reinforcement learning use a variety of different state representations to encode the system state, but most restrict themselves to one aspect of traffic (e.g. queue lengths, vehicle positions) and the current signal state. In this research, we propose a novel method of representing the state space through an abstracted bit-level matrix. A representation of this matrix is shown in Figure 3. In this matrix, we use various columns of the matrix to represent different components of the state space, including queue length, signal state, and time of day.

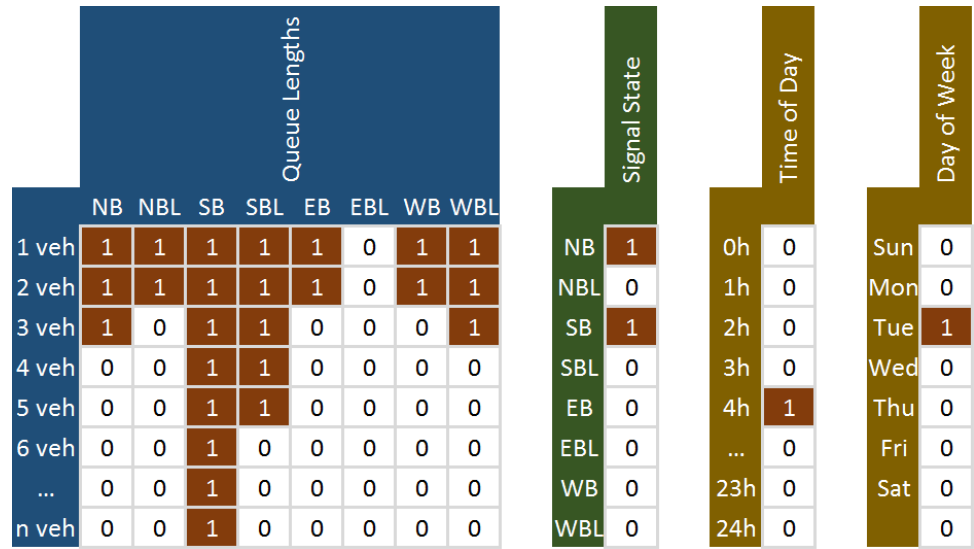

**FIGURE 3 – State Space Representation**

Queue lengths are depicted as abstracted columns of bits, with each individual cell in the column corresponding to an additional vehicle that is not moving at that intersection. The signal state is represented as a single column where various rows correspond to which signal state is timing. The time of day and day of the week is similarly encoded. The resulting state space is a matrix that can take be represented in a form similar to popularly used QR bar-codes.

Although still not in widespread deployment, video detectors have seen extensive piloting as technologies that can replace loop detectors. Video cameras have a number of advantages when compared to traditional loop detectors, including the ability to potentially supply more detailed information on the traffic state. Systems able to detect queue lengths have been proposed and field tested in the literature (*23*), though these systems are not without their limitations, including decreased accuracy at night and during inclement weather. Implementation of the traffic control strategy proposed here is therefore designed take advantage of technologies such as these, but could also theoretically function with loop detector based solutions.

*Model Structure*

By using an abstract representation of the system state, the importance of each component in the decision making process is left to the agent to determine. State spaces are made available at each second. Rewards are framed on the basis of discharged vehicles, with each vehicle incurring a reward of 20 utility for each vehicle it discharges, and a penalty of 1 disutility for each vehicle that is waiting behind a red light. Furthermore, an additional 5 disutility is added to each vehicle in a queue that is not fully discharged during a green cycle. Note that this rewarding scheme can be easily customized to account for different types of vehicles, and in this evaluation these values were selected in reference to existing literature in this area. In the future, a sensitivity analysis on the effect of this parameter would be beneficial as it likely has effects on the convergence speed of the model. Vehicles with speeds below 15 kph were considered as "queuing" for the purposes of both the rewards and the state representation. The state space used in this study allows application of convolutional neural networks, which have seen substantial success in many other pattern recognition problems, such as image recognition. These models work by applying successive layers of filters to the input, and typically have only a few weights. They act as a window, creating output for the next layer by scanning across the input. A key parameter in their configuration is "stride", which represents how "fast" the layer scans the input (e.g. how far the "window" moves when scanning). This parameter determines the size of the output from the convolutional layer, and if set to a value greater than 1 results in down-sampling. By applying more layers, the size of the input into the next layer is gradually reduced, with the goal being identification of key patterns. During training, the weights in these layers are adjusted incrementally as the system identifies patterns between the input states and the consequences it receives from its actions.

Figure 4 below shows the overall structure of the model proposed. Three convolutional layers are used to filter the input, and a pooling layer is inserted between the first and second convolutional layers.

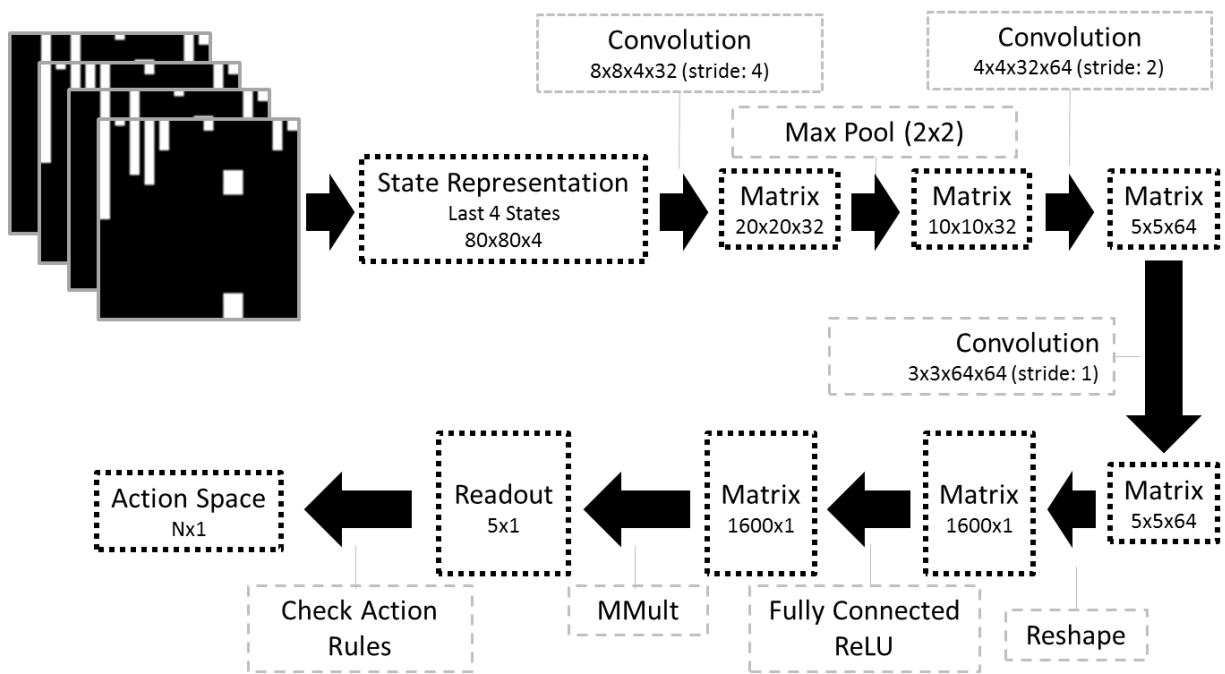

**FIGURE 4 – State Space Representation**

Inputs are framed on the basis of the last 4 states, and the state space matrix is 80x80. The input was reshaped to fit into the square matrix, and the individual columns were expanded to span multiple columns of the state space. With 80 rows, queue lengths in excess of 80 are not distinguished in the matrix. The remaining columns had row values calculated such that the entire column was used.

*Action Space Definition*

Although the system is designed to handle more complex situations, the examples trained in this study are limited to two actions. At each second, the controller decides between continuing the current phase, or commencing the next phase. The system is designed to roughly follow a ring barrier plan similar to how modern actuated signals function. Instead of relying on the concept of calls to guarantee the selection of a phase, the system can take an additional two actions (which are not used in this study, and thus never selected), including advance the first ring to the next phase only, advance the second ring to the next phase only, advance both rings to the next phase, or advance both rings directly to the phase before the next barrier. A rule-checking module determines which actions the controller is able to choose from at any given time, ensuring that only valid combinations are available. Additionally, after choosing to advance the signal, a minimum green time of 10 seconds is enforced, and the controller is not able to select actions until this time has passed other than the "do nothing" action.

The approach taken with the action space strikes a balance between flexibility and ensuring the controller does not choose counter-productive actions.

**RESULTS**

*Training Speed*

The model was trained using an epsilon-greedy process, which divides training into three stages. During the first stage, commonly called "observation", the model chooses the action with the highest expected Q value with probability $(1 - \varepsilon)$, and chooses a random action with probability $\varepsilon$. Here $\varepsilon$ is some initial value between 0 and 1, with 1 corresponding to fully random actions. Then, during the second stage, which is called "exploration", epsilon is annealed from its initial value to some final value $\varepsilon_f$. Annealing is normally done linearly over a defined time period. Finally, a "training" period commences where the model chooses the optimal action with probability $(1 - \varepsilon_f)$. $\varepsilon_f$ is normally not set to zero during the training process, as permitting the occasional random action allows the system to prevent itself from being trapped in a local minimum.

61 days of simulated data were used to train the model, and a final 2 days of simulated data were used for a comparative analysis. For each of the training days, a different random seed was used to generate the volumes and flows in VISSIM. The observation phase was designed to take place over the first 1.5 days, and the exploration phase over another 1.5 days for a combined total of 3 days. The initial $\varepsilon$ was set to 1, and the final epsilon to 0.005. At each second, the agent chooses an action, for a theoretical total of 86400 possible action choices in a day. During the training stages, random actions are selected from available actions with equal probability. To allow the model to experience a variety of states, during the observation and exploration stage, if the "do nothing" action is chosen randomly, the system restricts the model from choosing further actions for a randomly determined period of ranging from 0 to 15s. Subsequent random choices of the "do nothing" action during this phase still trigger this, and thus inactive periods longer than 15s may be experienced by the model.

Training time is visualized as the model's rate of improvement with each simulated day. Figure 5 plots the total network travel time across the simulated days. The Observe and Explore phases are highlighted in the figure. As can be seen from the figure, after 4 days of training the improvements in network performance in terms of total travel time begins to level off. Over the next 15 days, the system fluctuates and occasionally makes mistakes that result in large delays. Oscillations in the performance of the system begin to fluctuate less and less as the deep learning model learns the optimal control policies.

To analyze the effect of the model structure on the system, an alternative model was framed with a reduced state space (24x24) and with no pooling layer between the first and second convolutional layer. The first convolutional layer was also reduced in size from 8x8 to 6x6 and had its stride set to 2. The model was trained over 20 simulated days and compared to the original model. The resulting performance comparison is also plotted in Figure 5.

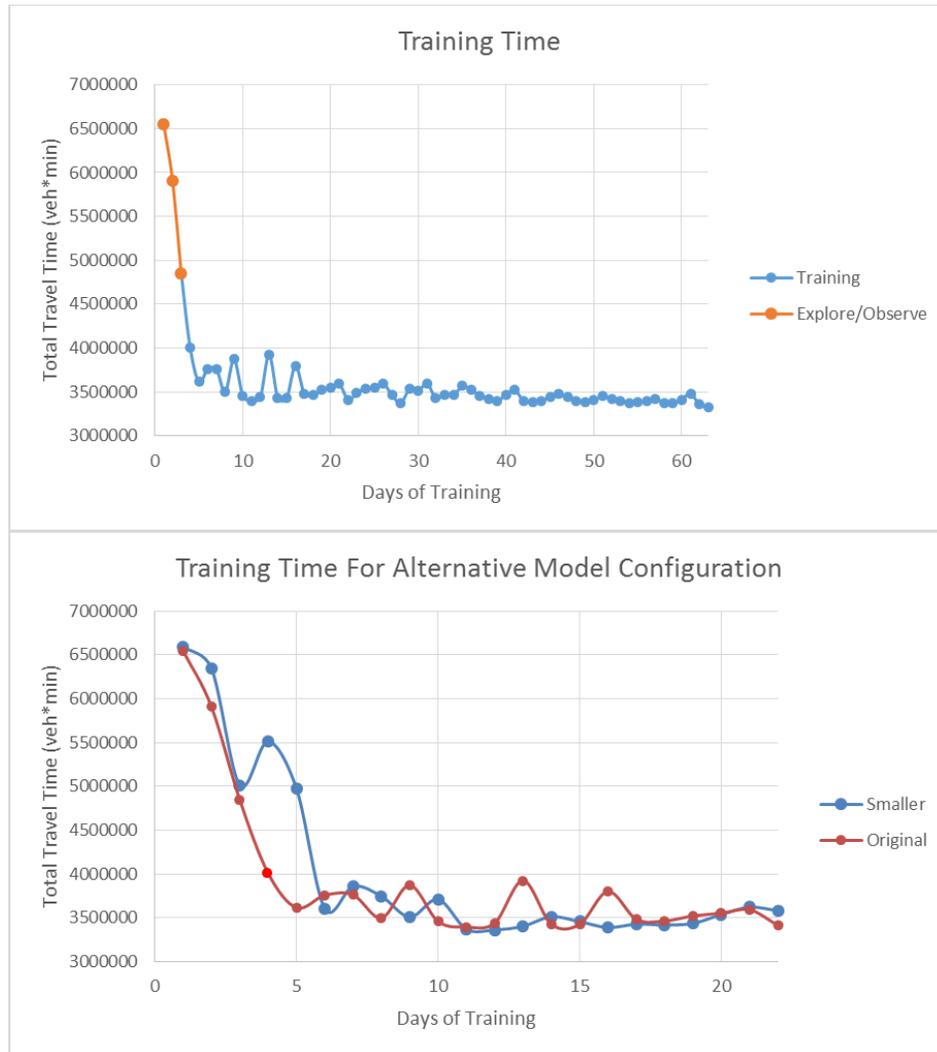

**FIGURE 5 – Total Travel Time Over Simulated Runs**

The use of the smaller model slightly reduced the initial learning speed of the model, but similar performance was observed after 6 days of training. This result suggests that a larger model may have a quicker learning speed, especially during the initial stages of training.

*Model Performance*

The proposed deep learning structure was evaluated by comparing it to a traditional semi-actuated and fixed time controller. For this configuration, the actuated controller has 4 timing plans it can rotate between, a morning peak, noon off-peak, afternoon peak, and an evening off-peak timing plan. For each timing plan, the time period with the highest volume was used to time signals in Synchro. Between 11 PM and 6 AM, the actuated signal operates in "Free" mode and uses no set cycle time. Synchro was configured according to the City of Toronto's Synchro guidelines, which are available online and are the parameters the City requires its consultants to use (*24*).

Results in VISSIM were obtained from a single day simulated on the trained deep learning model and traditional controllers. The same random seed was used in VISSIM for each method. Figure 6 below plots the average delay over the analysis period for the final day of the simulation.

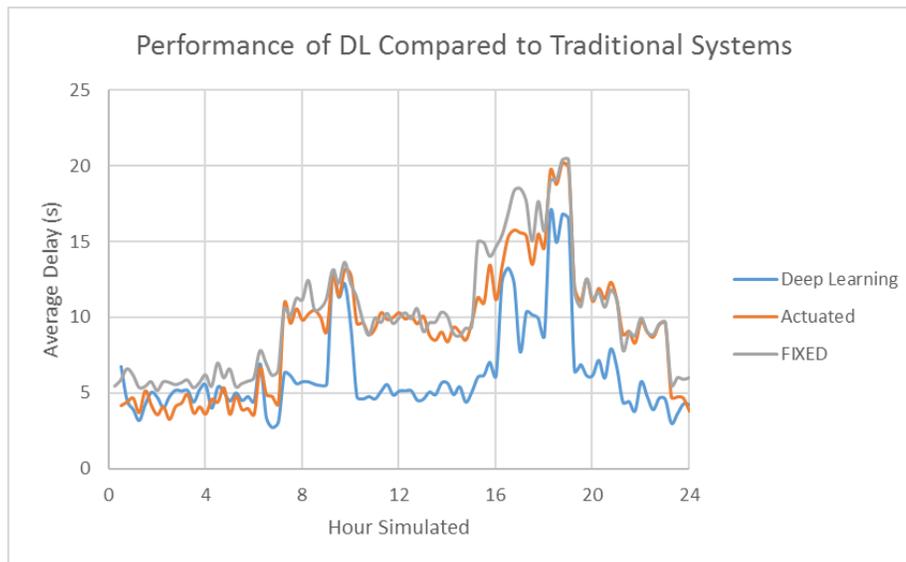

**FIGURE 6 – Average Delay, Performance of DL vs Traditional Controllers**

The benefits of the deep learning approach are most apparent in situations where the intersection is relatively under-saturated, namely during the "off-peak" periods. During peak periods operating at near-saturated conditions, the deep learning model performs similarly or better than the actuated controller. During periods of low demand, the deep learning controller and actuated controller perform similarly. Overall, the proposed deep learning controller reduces delays by 32% when compared to the actuated controller, and by 37% when compared to the fixed time controller.

**Unexpected Volume Scenario**

In order to evaluate the effectiveness of the deep learning model to rapidly changing traffic conditions, a hypothetical situation causing a spike in demand on one of the approaches was simulated in VISSIM. Such situations are not uncommon in large cities, as events such as major sport matches may disrupt normal flow patterns. The simulated scenario was designed to emulate the end of a major sporting event causing additional flows on the minor side street. Starting at 9 PM and ending at 11 PM, volume on the southbound approach of the minor street was set to 600 vph (from 175 vph in the training set). The performance of the deep learning model was compared to the actuated and fixed time controller, as in the previous scenario, with the results plotted in Figure 7. The period with the additional volume spike is visible on the figure as the region highlighted in red.

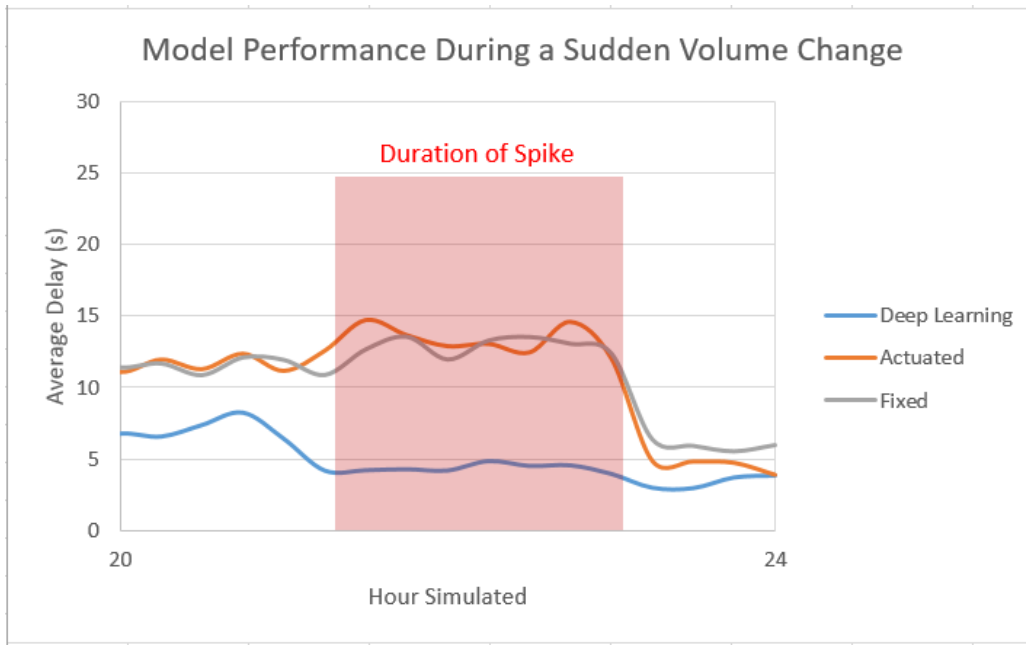

**FIGURE 7 – Average Delay, Performance of DL vs Traditional Controllers**

The situation modelled in this situation is still one where the minor approach is not over-saturated. Despite this, network-wide delay is higher in the actuated and fixed time scenarios as vehicles on the side street must wait longer for a green light. Although the training data set does not feature any situations where the load on the side street is similar to the main street, the deep learning model reacts to the situation by reducing the green served to the main corridor.

**CONCLUSIONS AND LIMITATIONS**

The results of analysis conducted in this research highlights the potential for deep learning and traffic signal control. Evaluation of the proposed system showed substantial improvements when compared to fixed time and actuated controllers under most conditions, but similar performance was observed in situations under-saturated off-peak conditions with very low volumes. Overall, the proposed system reduced average delays by 32% when compared to a semi-actuated controller and 37% when compared to a fixed time controller.

In this research we propose a novel way of framing the state space that abstracts various elements of the environment, including performance measures like queue length, signal state, and time of day into a single matrix. This approach is largely insensitive to the particular layout of an intersection, which means a model pre-trained on a wide variety of conditions may be able to respond to conditions so long as valid input is provided to it.

The results of this analysis also highlighted that the deep learning system is able to adapt to unexpected changes in volume which cannot be considered in traditional control methods without special timing plans. The result highlights the promise of the approach as an adaptive system. For reinforcement models, training speed may be a concern as it is often not possible to train a model on all situations. In this instance, volume on one approach was nearly four times its regular volume, but the model was still able to react appropriately.

*Limitations and Future Research*

The results of this study highlight the potential for deep reinforcement learning models for traffic signal control, and a number of unanswered questions still remain. Although the model achieves reasonable performance after a short amount of training, during the first 30 days of training, the model would occasionally make a suboptimal set of actions resulting in a major delay.

A number of other questions related to the model structure still remain unanswered. In particular, while models of different size were compared in this study, a comprehensive comparison is still required to fully ascertain what structure is optimal. The speed at which the model learns new features is important, as this affects its ability to adapt to new situations quickly, such as using a pre-trained model and deploying it at different intersections.

Finally, the proposed model only considers the simple case of an isolated intersection with two phases. The model is designed to handle more complex situations and evaluation for these cases is planned. Future research will also make use of deeper model structures and alternative state space framings.

## ACKNOWLEDGMENT

This research was supported through a project funded by NSERC (National Science and Engineering Research Council).